\documentclass{mn2e}
\usepackage{epsfig}
\usepackage{pslatex}

\def\msun{{\rm M_{\odot}}}
\def\rsun{{\rm R_{\odot}}}

\def\today{\number\year \ \ifcase\month\or
  January\or February\or March\or April\or May\or June\or
  July\or August\or September\or October\or November\or December
 \fi \ \number\day }
\date{Accepted ??. Received ??; in original form \today}

\volume{000}

\pagerange{\pageref{firstpage}--\pageref{lastpage}} \pubyear{2004}

\begin{document}

\label{firstpage}

\title[Young stars in the Galactic Centre]{Young stars in the Galactic Centre: a potential intermediate-mass star origin}

\author[L.\,M. ~Dray, A.\,R. ~King and  M.\,B. ~Davies]{L.\,M. ~Dray$^{1}$\thanks{E-mail: Lynnette.Dray@astro.le.ac.uk}, 
 A.\,R.~King$^{1}$ and M.\,B. ~Davies$^{2}$\\ 
1. Theoretical Astrophysics Group, University of Leicester, Leicester, LE1~7RH, UK\\
2. Lund Observatory, Box 43, SE-221 00 Lund, Sweden \\
}

\maketitle

\begin{abstract}
There has been recent speculation (Davies \& King 2005) that the cores of intermediate-mass 
stars stripped of their envelopes by tidal interaction with the supermassive black hole in the 
Galactic centre could form a population observationally similar to the so-called Sgr A$^{*}$ 
cluster or `S' stars, which 
have close eccentric orbits around the hole. We model the evolution of such stars, and show 
that the more luminous end of the population may indeed appear similar to young B stars  
within the observational limits of the Galactic Centre region. Whether some or all of these cluster stars can 
be accounted for in this manner depends strongly on the assumed IMF of the loss 
cone stars and the scattering rate. If most of the observed stars are in fact scattered from
the Galactic Centre inner cusp region itself then the population of $\sim 20$ to 
current observational limits may be reproduced. However, this only works if the local 
relaxation time is small and relies on the cusp stars themselves being young, i.e. it is 
dependent on some star formation being possible in the central few parsecs. 
Conversely, we obtain a possible constraint on the tidal stripping 
rate of `normal'-IMF stars if there are not to be red stars visible in the Sgr A$^{*}$ cluster. 

\end{abstract}

\begin{keywords}
Galaxy: centre -- Galaxy: stellar content -- stars: early-type -- stars: kinematics 
\end{keywords}

\section{Introduction}

In recent years, our ability to make observations of the region around the central black hole of our Galaxy 
has significantly improved. What we see there, however, remains somewhat of a paradox. It is becoming apparent that 
the stars we observe closest to the centre are predominantly hot, blue stars -- even though relatively unobscured 
observations can only be made in the H and K bands which are strongly biased towards picking up those stars which are 
redder -- and that their appearance suggests that they are young and massive. Genzel et al. (2003) observe two 
disks of hot stars within the central 1'' -- 10'' (0.04 -- 0.4 pc), which appear to contain abnormally high numbers of 
Wolf-Rayet (WR) and luminous blue variable (LBV) stars (Paumard et al. 2001), both products of the later stages of 
massive stellar evolution. There is 
some evidence that both this region and the Galactic Centre area in general have initial mass functions (IMFs) 
which are either biased towards 
the high-mass end or else have an abnormally high lower mass limit with respect to a standard Salpeter-like IMF (Nayakshin 
\& Sunyaev 2005, Stolte et al. 2005). However, the extreme tidal forces in the Galactic centre region make the normal formation 
of stars {\it in situ} a difficult task. Either these apparently massive stars must have formed in a non-standard manner 
or else they must have migrated inwards in a timescale short enough to accommodate 
the lifespan of an O star (Hansen \& Milosavljec 2003), a few $10^{7}$ years at most. 

The Sgr A$^{*}$ cluster is located even closer to the central black hole than the hot star populations described above -- 
it contains the stars with the smallest known distances of approach to Sgr A$^{*}$ (Ghez et al. 2003). Unlike the 
disk stars, their orbits are isotropically distributed and highly eccentric ($e \sim 0.3$ -- $0.9$), with peribothra as  
low as 0.5 mpc. These stars are also blue. K-band spectroscopy (Ghez et al. 2003; Eisenhauer et al. 2005) 
suggests that they are young O and/or B stars. The formation problems associated with the massive stellar disks are 
even worse for the Sgr A$^{*}$ cluster. If these stars formed {\it in situ}, their formation was subject to even 
stronger tidal forces. Similarly, any inward migration is likely to have been even more strongly impeded. It has been 
suggested that they are the remains of a shredded open cluster drawn inwards by dynamical friction 
(Kim \& Morris 2003), perhaps containing an intermediate mass black hole to keep it tightly bound 
enough to get in close enough to deliver them to their current location 
(McMillan \& Portegies Zwart 2003). However, these solutions would lead to spatial distributions of stars 
different from those observed (Ghez et al. 2005). Perhaps the most promising solution is that the high-velocity stars (HVSs) 
observed to be heading away from the Galactic Centre with velocities of hundreds of ${\rm km\,s^{-1}}$ (e.g. Brown et al 2005) 
share a common origin with the 
Sgr A$^{*}$ cluster, via the splitting of binary systems which encounter the black hole. In this scenario, one component 
of the binary is ejected to become a HVS and the other is captured (Gould \& Quillen 2003; Perets, Hopman \& Alexander 2006).

One possible solution to the problem is that the Sgr A$^{*}$ cluster stars are not young stars at all -- that they 
merely look like them. As proposed by Davies \& King (2005), another way to make blue stars in the Galactic centre region 
is to strip away the outer layers of giants, exposing their hot cores.
On many of the orbits on which the Sgr A$^{*}$ cluster stars currently reside, a RGB or AGB star would overfill its 
Roche lobe by several times at the point of closest approach to the black hole. Given that another of the well-known 
mysteries regarding the central region is the lack of red giants therein (Genzel et al. 1996; 
but see also Jimenez et al. 2006), this suggests that 
any giant star which ends up in the region after scattering onto an orbit passing close to Sgr A$^{*}$ quickly 
loses some or all of its envelope through tidal interactions with the black hole.
 The exposed core of such a star will appear blue, and it is possible under 
the restricted observing conditions imposed by the Galactic centre 
environment that it may resemble an O or B star for the portion of its stripped lifetime that it is visible.

This formation route for the Sgr A$^{*}$ cluster stars also accounts for their observed tightly-bound, 
eccentric orbits. Although the scattered star initially approaches the black hole on a near-parabolic orbit, 
the energy required to unbind the envelope is of a similar order to the orbital energy loss required to 
produce orbits similar to those observed. This energy requirement also puts constraints on which progenitor 
stars can feasibly reproduce the observed Sgr A$^{*}$ cluster population. Input stars must be able to 
survive the stripping process without being completely shredded.
Those for which the stripping of the entire envelope does not provide enough 
energy loss to produce a closely-bound orbit will end up on nearly-parabolic orbits in which they are only  
observable close to pericentre -- a negligible fraction of their lifetimes. Input stars must also have large 
radii to end up on orbits similar to those observed. The initial distance of closest approach to the black 
hole corresponds to the periastron distance in the post-stripping orbit. At periastron distances similar to 
those observed, a main-sequence star is not large enough to fill its instantaneous tidal lobe. Such stars have to pass 
far closer to the central black hole to undergo tidal mass loss. This, combined with their less centrally-condensed 
structures, suggests that main-sequence stars are probably shredded completely once they have filled their tidal lobes.
Envelope stripping in the central region, albeit by stellar collisions, has also been invoked by Alexander (1999) to explain
the distribution of stars in the inner 0.05 pc. 

There are further observational constraints that these stripped giants must obey if they are to be a plausible 
explanation for the Sgr A$^{*}$ cluster. The cluster stars all appear to be on or near the main sequence; therefore we 
require that the observable population of stripped stars must also primarily reside in this area of the HR diagram. 
Since K-band 
visibility increases rapidly with decreasing ${\rm T_{eff}}$, this is (if one ignores the potential effect of low-luminosity AGN 
irradiation on their atmospheres; Jimenez et al. 2006) a strong constraint on how red the stars may be. However, a 
population bluer than the main sequence is simply invisible in the K-band to current observational limits and is therefore 
only constrained by the requirement that the scattering event rate (Wang \& Merritt 2004) and the mass of unobservable remnants 
(Goodman \& Paczynski 2005) be realistic. The stripped stars must also display a similar spectrum in the K-band to those observed. 
This requires, for instance, a composition including a non-negligible surface hydrogen abundance, and an 
absence of Wolf-Rayet-like emission lines.  Lastly, they must also be capable of reasonably rapid rotation; for 
example, the star S2 (S0-2) has a $v \sin i$ value of around $220 {\rm km\,s^{-1}}$. Of these, the event rate is 
the most pressing problem. In this paper we consider whether those 
constraints can be met by running simulations of rapidly-stripped stars from different input populations and comparing them with 
the observations.

\section{Simulations}

The process of tidal stripping is a complex one, and it is far from certain that the apparent balance between 
envelope binding energy and orbital energy translates into the possibility of a real and efficient 
transfer of energy between the two. To study this, we take a two-pronged approach. Firstly, the tidal mass loss 
itself and its effect on the orbit is being studied by means of detailed SPH simulations (Dale et al., in preparation). 
Secondly, the evolution of stripped stars with simple assumptions about the stripping process over a range of input parameters
is calculated in order that their properties may be compared with those of the Sgr A$^{*}$ cluster stars. This is the subject of the 
current paper. In doing so we have assumed the stripping process itself and the uncertainties relating to it (whether, 
for example, there is really an efficient conversion between orbital and envelope binding energy, and whether the stellar 
core is able to survive the process) to be a `black box' to be explained or disproved by the SPH simulations. 

\subsection{Code}

We use the Eggleton stellar evolution code (Eggleton 1971, Pols et al. 1998), including the convective overshooting
prescription of Pols et al. (1998) with $\delta_{\rm ov} = 0.12$, which is roughly equivalent to an overshooting distance of 
$0.2 {\rm H_{p}}$, where $H_{p}$ is the pressure scale height. This leads in general to relatively large model cores.
All of the runs in this paper were carried out at 
a metallicity of ${\rm Z = 0.02}$ as appropriate for the Galactic Centre region (Najarro et al. 2004; Carr, Sellgren 
\& Balachandran 2000). A slightly higher or lower metallicity is unlikely to make a strong difference.
Wind mass loss is included according to the prescription of de Jager et al. (1988), but is 
generally small for most of the masses we have considered. Whilst we can run the code following binary 
evolution including mass loss to a companion (in this case the black hole), 
we here treat our star as a single star, and a constant high rate of 
mass loss to it at the appropriate age to simulate the stripping event. 

\begin{figure}
\vbox to142mm{\vfil
\psfig{figure=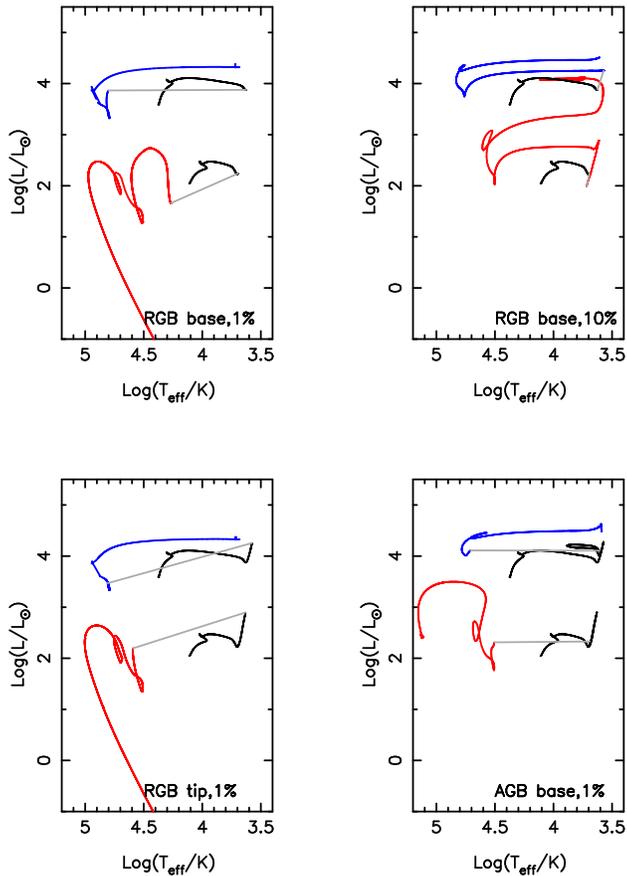,angle=0,width=82mm}
\caption{Typical evolutionary tracks for intermediate mass stripped stars pre- and post-stripping. Shown in each case are a $3.25 \,\msun$ (lower track) and an $8.85 \,\msun$ model (upper track) Pre-stripping tracks are shown in black, the rapid mass loss and subsequent adjustment phases are indicated by the grey lines and post-stripping tracks are shown in blue ($8.85 \,\msun$) and red ($3.25 \,\msun$). The stage at which stripping occurs and the amount of envelope remaining are given on each panel.}
\vfil}
\label{fig1}
\end{figure}

The rapid nature of the mass loss from the stars we are considering presents a problem for 
simulation. Essentially, the entire envelope may be removed in a single orbit of the black hole; 
this equates to a mass-loss rate which may be over a solar mass a year, which is difficult to 
apply in any realistic way to a stellar evolution code. What we can do, however, is apply mass loss at the fastest rate which 
the code will sustain. This is an approximate approach, as equilibrium is assumed throughout, 
but gives a general idea of how these stars will behave. The mass-loss 
rate used varies slightly from star to star -- since we 
use in each case the largest rate the code will support without breaking down -- but is generally between $10^{-1}$ 
and $10^{-3} \msun\,{\rm yr}^{-1}$. 
In all cases the entire mass loss event lasts significantly less long than would be a typical timestep for 
normal evolution of that star at that evolutionary stage. From the evolution of other stars 
which are or have been affected by rapid mass loss -- the primaries in interacting binary systems, 
subdwarf O and B stars and massive Wolf-Rayet stars -- it is possible to infer that the most likely course 
for our stripped stars is a rapid journey to the blue part of the HR diagram followed by a relatively 
stable period of helium-burning on the He main sequence, as can be seen in Fig. 1.

\subsection{Model Grid}

A number of variables will determine the evolution of a star after its envelope has been removed. 

\subsubsection{Initial Mass}

There is no absolute upper or lower mass limit on stars which interact with Sgr A$^{*}$. 
However, the range of plausible pre-stripping masses for at least the visible cluster stars is 
limited at the lower end by stars which are unlikely to become luminous enough to be 
observable in the Galactic centre region, and at the upper end by stars which are likely to have a Wolf Rayet-like appearance when 
stripped and are thus relatively easily distinguishable from B stars. We limit our detailed simulation input mass range to 
between 1 and $50 \msun$. In this range we run model grids at 21 log-spaced masses, and interpolate as required between the 
resulting tracks. However, we consider stars over the whole range of potential IMFs by using the synthetic stellar evolution 
formulae of Tout et al. (1997) for lower masses. 
For these stars it is highly unlikely that we can create a blue core which is as luminous as those observed -- 
a rapid transition to white dwarf status is the most likely outcome of stripping. Furthermore, most stars below $1 \msun$ 
have sufficiently long evolutionary timescales that at the current age of the Universe they will still be on the main sequence, and are 
therefore ruled out as progenitors for at least the currently-observed Sgr A$^{*}$ cluster stars.
Whilst it is possible that the Sgr A$^{*}$ cluster stars are 
stripped stars in the upper region of or even above the mass range we simulate, the short lifetimes of very massive 
stars (a few Myr) are a strong 
impediment to getting them in to the Galactic centre region in time. It is still possible, however, that they could have been 
scattered from the disks of stars at $\sim 0.4$pc, which are massive and may have been formed {\it in situ} 
(Nayakshin \& Sunyaev 2005). Normally intermediate-mass stars are strongly favoured by the IMF over 
more massive stars, but there is evidence that in the Galactic Centre region the IMF is significantly top-heavy 
or may have a high lower limit (Stolte et al. 2005). 
However in this case the original (albeit not the current) masses of the Sgr A$^{*}$ cluster stars would be 
greater than that usually inferred in the case that they are B stars! In this scenario the `Paradox of Youth' (Ghez et al. 2003) is 
not resolved; at the least it is a requirement that the massive star disks formed in their current location. Such an origin may also 
require that the relaxation time in the central parsec is shorter than the life of a $\sim 6 \msun$ star, i.e. $\sim 10^{8}$ years.

The simulations are ended either when the star has become a white dwarf, when degenerate carbon burning 
ignites or when central carbon burning ends for the more massive models. 

\subsubsection{Evolutionary Stage at Stripping}
Since tidal stripping can occur rapidly when a giant is scattered onto a orbit which passes close to the central black 
hole, the evolutionary point at which the stripping occurs does not necessarily correspond to the stages in the star's life 
when it is beginning to expand, as it would for Roche lobe overflow (RLOF) in a binary system. Instead, this violent RLOF-like event may occur at 
any point in the star's life when its radius is above the instantaneous tidal radius on the scattered orbit,
\begin{equation}
R_{\rm tidal} \simeq 0.462 \left(\frac{M_{*}}{M_{\rm BH}}\right)^{1/3}\,p \,\,,
\end{equation}
where $p$ is the periastron and $M_{*}$ and $M_{\rm BH}$ are the masses of the scattered star and the black hole respectively. 
However, analogously to RLOF, there is also the possibility that stars which are scattered onto near-parabolic orbits when their 
radii are too small for 
the stripping process to take place (as is likely for main sequence stars, which should form the bulk of the scattered population) 
may be stripped on a subsequent orbit of the black hole when they have evolved into larger-radius objects. This requires that 
the orbital time be less than the evolutionary timescale of the star -- in the case of our best-matching models 
a period of a few $10^{6}$ years is probably the maximum for these orbits, corresponding to an initial tidal interaction on first pass 
close to the black hole which extracts some $10^{46}$ ergs of energy from the orbit. Depending on the relaxation time in the region 
near Sgr A$^{*}$ and the resulting period, some of these stars are also likely to be scattered back off these long orbits before 
they can be stripped. 

     To allow for a distribution of scattering times, we simulate for each of our stellar models the onset of rapid mass 
loss at the RGB base, middle and tip, on the horizontal branch and at the base and middle of the AGB. 
Stars which approach close enough to tidally overflow whilst they are still on the main sequence are likely to be completely disrupted. 

\subsubsection{Amount of Envelope Remaining}
Without detailed calculations of the effect 
the tidal forces have on the star (currently in progress) it is hard to say exactly how much of the envelope will be stripped. For 
the vast majority of possible orbits the answer will be `nearly all' but the remaining amount can strongly affect how blue the star 
becomes, if it becomes blue at all, and also its remaining lifetime after the stripping event. In order to get an idea of 
how much of the envelope needs to be removed before a blue star is obtained, we run models at each of the above grid points which are left 
with 0, 1, 5, 10, 20 and 50 percent of their envelopes. The latter few situations are unlikely -- requiring 
scattering onto a grazing orbit and therefore relatively mild mass loss. Typical evolutionary tracks are shown in Fig. 1. 
Generally stars with more than about 5 percent remaining envelope spend a significant amount of their subsequent evolution  
as red giants. 

The amount of envelope left strongly affects the orbit the stripped star ends up on, as it limits 
the amount of energy which can be extracted by envelope removal. It is possible that different 
Sgr A$^{*}$ cluster stars have different amounts of remaining envelope. 
It is also notable that since the smaller the mass of the stripped core, the smaller its orbital 
energy (in an equivalent orbit), the criterion that the binding energy of the removed envelope be 
comparable to the orbital energy a typical Sgr A$^{*}$ cluster star would have if it were at the remaining 
core mass is fulfilled by a wide range of initial masses, provided the star is large enough to overflow its tidal lobe in that orbit.
This is illustrated in Fig. 2 for stellar models at various masses.

\begin{figure*}
\vbox to140mm{\vfil
\psfig{figure=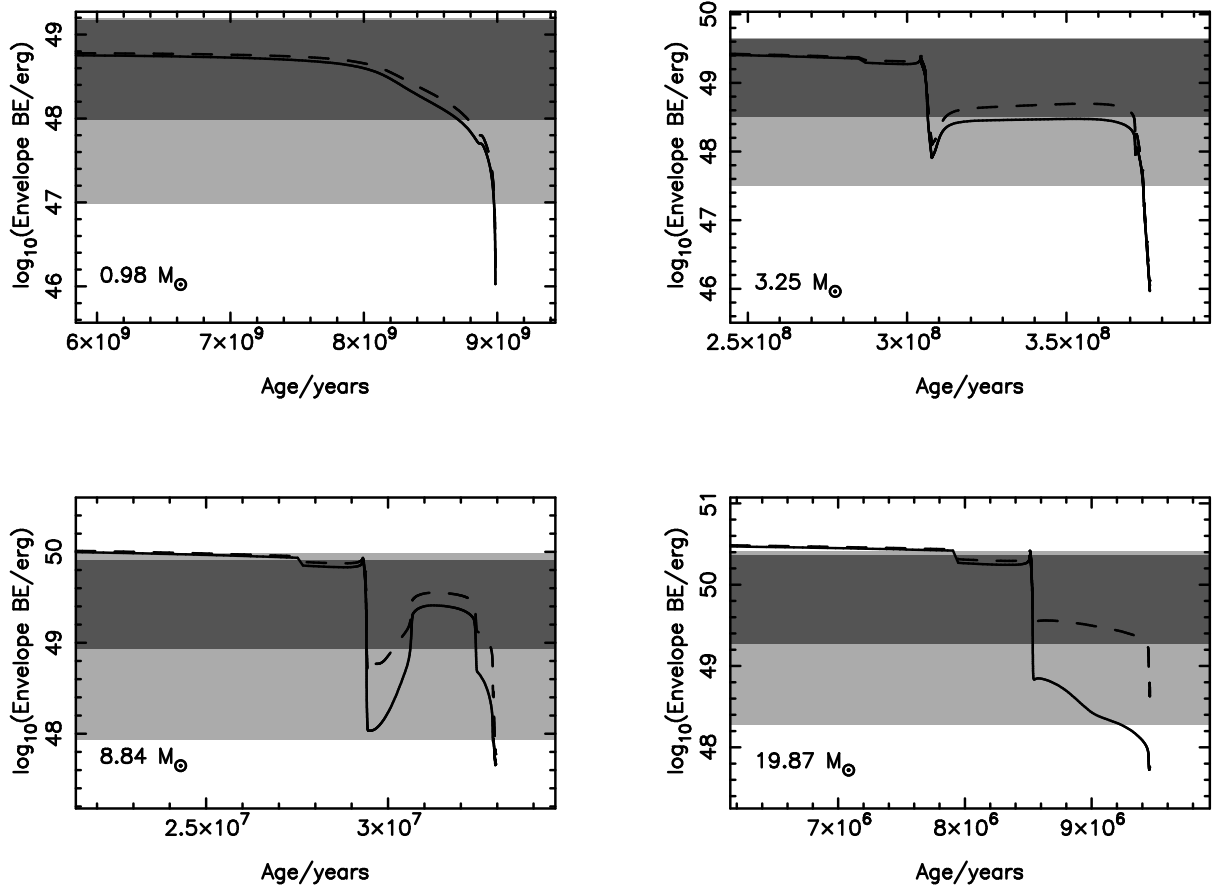,angle=270,width=160mm}
\caption{Energy required to strip 90 \% (solid lines) and 99 \% (dashed lines) of a star's envelope over the later phases of its 
lifetime, for different initial masses. Also shown greyed-out is the area in energy corresponding to the range of orbital energies which 
Sgr A$^{*}$ cluster stars with known orbits (Eisenhauer et al. 2005) would have if they were the mass of the resulting stripped stars. The darker grey area 
indicates the region of overlap between the energy range for the 90 \% and 99 \% stripped cores.}
\vfil}
\label{fig2}
\end{figure*}

\section{Input Population}

Simulations which calculate the disruption rate of stars by black holes frequently restrict the stellar input distribution to a single 
mass and radius (e.g. Wang \& Merritt 2004). However, that approach is clearly of little use here, since we wish to simulate a population 
which is likely to arise from a relatively small segment (the high-mass, large-radius end) of a diverse input population. Input masses will 
initially be distributed according to an IMF which, at least for stars originating from outside the immediate Galactic Centre region, 
is likely to be Salpeter-like (e.g. Kroupa 2001). However, the probability of a star ending up on an orbit which passes close to 
Sgr A$^{*}$ also depends on its mass and radius, as does its final orbit. 
If the star is not stripped on its first pass of the black hole, whether it can be stripped on a subsequent pass 
depends on its future radius evolution. 

\subsection{Scattering and event rate}
The tidal disruption loss cone of a galaxy is the set of orbits for which stars pass sufficiently close to the central black hole 
for tidal disruption to take place. This can only happen once per star so, as well as having to be populated in the first place, 
the loss cone must be continually refilled if the number of tidal disruptions is not to drop to zero with time (e.g. Merritt \& Wang 2005).
The total scattering rate for loss cone stars and the loss cone refilling rate have been estimated by various authors 
(e.g. Frank \& Rees 1976; Lightman \& Shapiro 1977; Rees 1988; Magorrian \& Tremaine 1999; Merritt \& Poon 2004; Wang \& Merritt 2004; 
Ivanov, Polnarev \& Saha 2005; Holley-Bockelmann \& Sigurdsson 2006) and may vary over several orders of magnitude depending on the 
assumptions used.
We use the scattering rate of Wang \& Merritt (2004), 
\begin{equation}
\dot{N} \approx 6.5 \times 10^{-4} \,{\rm yr^{-1}} \,\left(\frac{M_{\rm BH}}{10^{6} \msun}\right)^{-0.25} \, ,
\end{equation}
which utilises the M-$\sigma$ relation of Merritt \& Ferrarese (2001), and assume that $M_{\rm BH}$ is $3.61 \times 10^{6} \msun$ 
(Eisenhauer et al. 2005). Whilst the above rate is derived for solar-type stars only, it scales 
with mass and radius only as $m_{*}^{-1/3}r_{*}^{1/4}$. For the models we have run we find that the overall scattering rate integrated over 
realistic populations differs from this by less than a factor of 3, provided the slope of the upper IMF is Salpeter-like. 
The amount of difference depends on the IMF, with more top-heavy IMFs having smaller scattering rates. In the case of IMFs with flatter upper
slopes, such as that posited for the core of the Arches cluster (Stolte et al. 2005), the difference may be a factor of 10 or more.

 We assume stars, initially distributed according to either a Salpeter or top-heavy IMF, are supplied to the loss cone at a point in their life 
chosen randomly but weighted according to the mass and radius distribution above. This results in a number of distinctive characteristics  
for those stars which end up on loss cone orbits. Firstly, short lifetimes and the mass-dependence of the scattering rate 
ensure that few of the most massive stars ($> 15 \msun$)
contribute to the population, even for top-heavy IMFs.  Since such stars, once stripped, would probably display WR-like emission lines in their 
spectra unlike the observed Sgr A$^{*}$ cluster stars, 
this seems reasonable. Secondly, if one assumes a Salpeter IMF, most of the loss cone stars are in fact low mass main sequence stars which 
will be disrupted completely by a tidal-radius encounter with the black hole. Even if one assumes an IMF weighted towards more massive 
stars, as might be typical of the Galactic Centre region, most of the stars which are stripped will be main-sequence stars on close orbits. 
The Sgr A$^{*}$ cluster stars in this picture therefore arise from a relatively small proportion (a few percent) of the total number of stars 
scattered; those on the giant or horizontal branches.

A further complication arises in that tidal stripping is not the only possible outcome of stars passing close to the black hole. Stars which pass 
the black hole at a distance above a tidal radius may still dissipate energy due to tidal distortion of the envelope, becoming tidally captured 
but not stripped. Such interactions may occur for initial peribothron distances of about three tidal radii (Novikov, Pethick \& Polnarev 1992). 
On subsequent peribothra further tidal interactions take place, gradually causing the orbit to decay. If the star were able to survive these 
repeated interactions (which is unlikely, see Novikov et al. 1992) this would eventually result in a very close, circular orbit around the 
black hole. Relatedly, some main sequence stars are also scattered onto the red giant loss cone. On their first passage past the black hole these stars do not come 
close enough for significant tidal interaction at their current radii but they subsequently evolve to large enough radii
that stripping can occur on a later orbit (Syer \& Ulmer 1999).  It is possible that at least some of the loss cone orbits are chaotic 
and indeed that chaotic orbits may be required if the loss cone is to be kept full (e.g. Merritt \& Poon 2004, Holley-Bockelmann 
\& Sigurdsson 2006), 
in which case this population may be minimal. However, if not, the population of stars late-stripped in this way should be larger than the 
population from direct stripping on first peribothron passage. Since the radius-dependence of the scattering probability is relatively shallow 
the magnitude of this late-stripped stars effect is 
less than might at first be assumed. However, it is important in that nearly all the stars supplied to the black hole in this way are giants  
which are much less likely to suffer complete disruption than the main-sequence stars.  Therefore stars which are stripped on their first 
pass of the black hole give very much a lower limit to the number of potentially observable Sgr A$^{*}$ cluster stars. An estimate of 
the population from this later stripping comes from running our simulations with the assumption that a star is stripped if scattered 
onto the loss cone corresponding to its {\it maximum} rather than current radius, i.e. 
$P_{\rm scatter} \propto m_{*}^{-1/3}r_{\rm max,*}^{1/4}$. We then assign it a 
probability-weighted periastron distance in the range $r_{\rm tidal, current}$ to $r_{\rm tidal, max. radius}$ and add it into the stripped 
stars population when it is large enough to overflow at periastron. This approach overestimates the population only in cases 
where the evolutionary timescale is shorter than the orbital timescale, which is not the case for most stars. 
It should be noted that we restrict the maximum-radius calculation to ages of less than $10^{10}$ years, so as to exclude stars which cannot 
have evolved to their present state in the time available, e.g. $0.5 \msun$ giants. This leads effectively to a mildly top-heavy IMF 
with a turn-over at around $1 \msun$ amongst the stars which end up being stripped when a Salpeter-like IMF input population is assumed, 
since the maximum radii of more massive stars within the time limit are larger.

Given the r-dependence for stars passing within their tidal radius  
above and assuming stars typically increase their radii by a factor of a few hundred between the main sequence and giant phases, it seems 
likely that the late-stripping giant population is roughly a few times the population of main-sequence stars stripped on their 
first pass of the black hole. This is borne out by simulation results; the factor by which the scattered population increases varies from  
less than 0.1 for stars with a `normal' IMF (Kroupa 2001; the small change is due to the high proportion of low-mass stars) to over 3 
for stars with the most top-heavy IMF. However the top-heavy IMF populations had the lowest scattering rates to begin with.
We call models incorporating stripping after the first pass 'late-stripping' models.
As well as significantly increasing the visible Sgr A$^{*}$ cluster population and the number of non-visible remnants on similar 
orbits to the Sgr A$^{*}$ cluster stars in comparison to equivalent first-pass stripping models, adding in late-stripped giants 
can increase the accretion rate onto the black hole from tidal stripping  by a factor of a few. 
Since there can be a significant time delay associated with 
lower-mass stars reaching their maximum radii, the late-stripping Sgr A$^{*}$ cluster population is likely to 
change with age and will initially be much more top-heavy in mass than the input population.

\subsection{IMF}
Since the scattering probability varies inversely with mass, scattering from an initially Salpeter-IMF reservoir of 
stars leads to a significantly bottom-heavy mass distribution for stars that pass within a tidal radius of the black 
hole. This is compounded by the short lifetimes of the more massive stars.
However, it is those more massive stars which have sufficient luminosity to be observed in the GC region.
Of a `normal' Salpeter-like first-pass stripped population only around one star per 38000 is initially above $5 \msun$.  
With a scattering rate of $\sim 5 \times 10^{-4} \, {\rm stars \, yr^{-1}}$ this obviously produces a very poor 
fit to the observed population of around 20 luminous Sgr A$^{*}$ cluster stars\footnote{A $5$ -- $6 \msun$ star stripped on the RGB can live 
for 1 -- $2 \times 10^{7}$ years; since the scattering probability increases with radius, about a third of the scattered $> 5 \msun$ 
stars are post-main sequence. For about $5 \times 10^{5}$ -- $10^{6}$ years of this time they are observable in the 
K-band under conditions appropriate for the Galactic Centre. As stars below this mass spend shorter periods of their 
lives visible and blue, a rough calculation gives an Sgr A$^{*}$ cluster population over 300 times too small.}.

Though there remains the possibility that the scattering rate 
could be higher than the Wang \& Merritt one (e.g. Holley-Bockelmann \& Sigurdsson 2006), a large enough scattering rate 
would also violate other constraints imposed by observations of the GC region, requiring in particular an extremely large 
population of non-visible stellar remnants (Goodman \& Paczynski 2005; see also section 4.4.1). 

   There are two plausible effects which may help in increasing the Sgr A$^{*}$ cluster population without increasing the 
scattering rate, however. 
Firstly, many of the scattered stars come from the central region of the Galaxy itself. The simulations of Merritt \& Szell (2006) 
suggest a lower limit of $10^{-4}$ stellar tidal disruptions per year for stars from the central cusp region, i.e. the inner few parsecs. 
There are numerous 
indications that the GC region may have a top-heavy IMF (Nayakshin \& Sunyaev 2005). The present-day MF of the Arches cluster (around 30pc 
in projection from the central black hole) may have a turn-over at masses as large as $6 \msun$ (Stolte et al. 2005). Top-heavy IMFs 
with lower mass limits of $1 \msun$ or more can be a natural consequence of star formation in dense gas with a high opacity, such as 
in the GC region (Larson 2006).

Secondly, as discussed above, adding in a population 
of late-scattered stars favours more massive, faster-evolving stars which reach larger radii.

   As well as running our simulations with a `normal' IMF (we use the IMF given in Kroupa (2001), which has a Salpeter-like slope for 
massive stars and a lower limit of $0.01 \msun$),
we also run simulations with input IMFs which have lower mass limits at $1 \msun$ (leading to one in 2500 stars above $5 \msun$), 
$3 \msun$ (one in 7 stars above $5 \msun$) and $6 \msun$, and a hybrid IMF which assumes a rate of $10^{-4} \,{\rm disruptions \, yr^{-1}}$ from a GC IMF with a 
turnover at $6 \msun$ and the remaining $4 \times 10^{-4} \,{\rm disruptions \, yr^{-1}}$ from a normal IMF. For the $3 \msun$ lower 
limit IMF we also run models with a relatively flat slope for stars above $6 \msun$, as is posited for the Arches cluster.

\begin{figure}
\vbox to120mm{\vfil
\psfig{figure=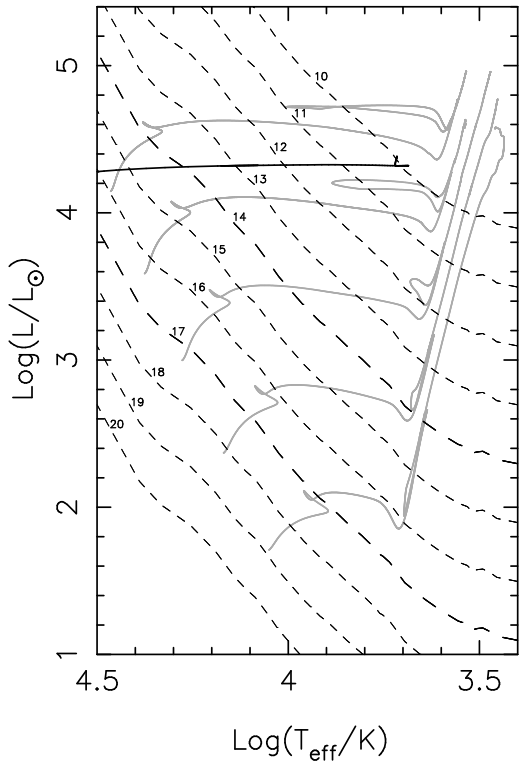,angle=0,width=68mm}
\caption{HR diagram showing K-band contours (dashed lines) from K = 10 to 20. The observed Sgr A$^{*}$ cluster stars lie between 14 and 16.8 (Eisenhauer et al. 2005).
Also shown are unstripped stellar tracks for 2.7, 4, 5.9, 8.8 and 13.3 $\msun$ stars (grey) and, for reference, a portion of the stripped track 
for the 8.8 $\msun$ RGB base star shown in figure 1 (black solid line).}
\vfil}
\label{fig3}
\end{figure}

Whether the top-heavy IMF models are realistic or not depends on the relaxation time in the central region, since these massive stars still have 
to get on to the loss cone. The lives of more massive stars are short. If the relaxation time is much greater than the main-sequence lifetime of 
potential Sgr A$^{*}$ cluster-star progenitors, then it is unlikely that enough of them are scattered. In particular, if the relaxation time is $10^{9}$ years 
then this suggests very little tidal stripping is possible from our top-heavy IMFs. Realistically, we require either a relaxation time below 
$10^{8}$ years or else that the current spatial distribution of stars in the central region (in particular the two inclined disks rotating in 
opposite directions) has some effect on the rate of stars scattering towards the black hole. The high scattering rate from the cusp region 
calculated by Merritt \& Szell only includes the contribution of stars at $r < 0.7$ pc, so the relaxation time of this region is appropriate here.  
Sch{\"o}del et al. (2003) suggest the high 
stellar density in the central arcsecond should lead to a relaxation time of less than $10^{8}$ years, 
and this value may be reduced further if one takes into account the potentially-large population of compact remnants which are likely to also exist 
there\footnote{Note though that if the density of objects in this region is too high then repeated interactions could strip some of the envelope 
from input stars before their interaction with the black hole. This would reduce the orbital energy budget and skew the resulting orbit distribution
towards higher eccentricities.}
 (Davies \& King 2005) and the possible effects of resonant relaxation (Rauch \& Tremaine 1996; Hopman \& Alexander 2006). A lower limit 
slightly further out is provided by the unrelaxed stellar disks at $\sim 0.4$ pc. However, the presence of Wolf-Rayet stars and LBVs suggests 
they are only a few $10^{6}$ years old. 
The top-heavy IMF scenarios are therefore at least not ruled out. If on the other hand they are not possible, 
then the alternative (given the evidence for a top-heavy IMF in the central region) is not lower-mass star scattering from the central region but 
a reduced overall rate of scattering in which all the loss cone stars come from much further-away regions with more normal IMFs.

\section{Observability}
    Observations of stars in the Galactic Centre are only possible in the infrared due to the 
heavy visual extinction (some 35 magnitudes) of the area. This presents particular problems for hot, blue 
stars for which most of the energy output is far from the observable region of the spectrum. Nevertheless, 
H and K-band spectra have been obtained for some of the Sgr A$^{*}$ cluster stars (Ghez et al. 2003, Eisenhauer et al. 2005).
For the brightest, S0-2 (S2 in Eisenhauer at el. 2005), Ghez et al. (2003)  observe the H I (4 -- 7) Br$\gamma$ line 
at 2.166$\,\mu$m and the He I triplet at 2.11$\,\mu$m; other lines in the region, such as N III at 2.1155$\,\mu$m, 
are not seen. They obtain an effective temperature for the star of around $30,000\,$K.
Using the hot star spectral atlas of Hanson, Conti \& Rieke (1996), this led them to conclude 
that S0-2 was a late-type O star. Further observations by Eisenhauer et al. (2005) confirmed that the majority of 
the Sgr A$^{*}$ cluster stars have B star-like spectra. Interestingly, they appear to fall into strong-lined and weak-lined groups, 
suggesting some form of dual origin or dual evolution is going on. 

However, for many types of massive star there is a significant amount of overlap in IR spectral morphology (Morris et al. 1996) 
leading to potential confusions in classification. 
Whilst it is difficult to find template spectra for stars which have undergone similar evolution to our putative Sgr A$^{*}$ cluster-forming 
process, to a first approximation they will appear similar to B stars (R. Napiwiotzki, private communication). Those stars 
which fall into the observable region shown above have surface gravities, luminosities and effective temperatures in ranges which are 
similar to B stars. One important constraint imposed by the spectra is that a significant surface hydrogen abundance must remain. 
However, whilst the stripped stars which remain luminous enough to be observed display He-enriched surface abundances 
after envelope stripping, nearly all retain a surface H mass fraction of between $0.1$ and $0.6$. Those which retain more envelope 
have higher surface H abundances; stars which lose their entire envelopes have, of course, rather low surface H ($< 0.12$ by mass fraction).

 To gain a rough idea of which stars will be visible, we calculate K-band limits assuming a distance modulus to the 
Galactic Centre of 14.6. For the K-band extinction we use the average value found by Blum, Sellgren \& DePoy (1996), $A_{\rm K} \sim 3.3$. 
However the authors note that $A_{\rm K}$ is rather variable, and can reach $\sim 6$ in places. Tabulated bolometric corrections and 
$V - K$ colours are taken from Johnson (1966). The resulting K contours are shown in Fig. 3. It is apparent that red stars in the galactic 
centre region are significantly more observable than blue stars. 

\begin{figure}
\vbox to140mm{\vfil
\psfig{figure=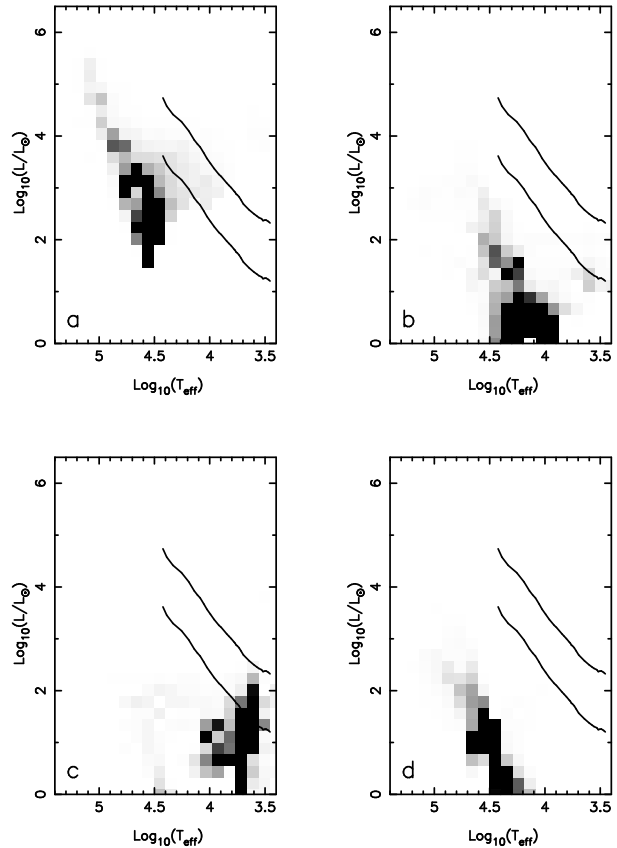,angle=0,width=80mm}
\caption{Density distributions of overall Sgr A$^{*}$ cluster star HR diagram positions for varying assumptions about envelope stripping and IMF. Panels {\bf a} and
{\bf b} show populations with 1\% of envelope remaining and top-heavy (turnover at $6 \msun$) and `normal' Salpeter-like IMFs respectively. 
Panels {\bf c}
and {\bf d} show the Salpeter IMF with 5\% and 0\% remaining envelope. Shown also are the approximate equivalents of the upper and lower 
K-band luminosity limits observed for the Sgr A$^{*}$ cluster stars.}
\vfil}
\label{fig4}
\end{figure}

In Fig. 4 we show the effect of IMF and amount of envelope remaining on the HR diagram positions of the stripped stars. It is apparent, as 
noted previously, that the `normal' IMF populations cannot reproduce a luminous-enough group of stars. In addition, unless nearly all 
of the envelope is removed, they produce 
a population of relatively low-luminosity red stars which will be visible in the K-band. It is thus unlikely that a normal-IMF first-pass population
will be able to reproduce the Sgr A$^{*}$ cluster stars even if the scattering rate is higher than previously assumed. However, the populations with top-heavy IMFs 
produce a number of luminous blue stars, although many of these populations also produce some red stars as well. For simulations including only 
first-pass stripping the visible populations are generally blue, but insufficiently numerous. This can be seen in 
snapshots taken at points during the simulations, as in Fig. 5. 

\begin{figure}
\vbox to160mm{\vfil
\psfig{figure=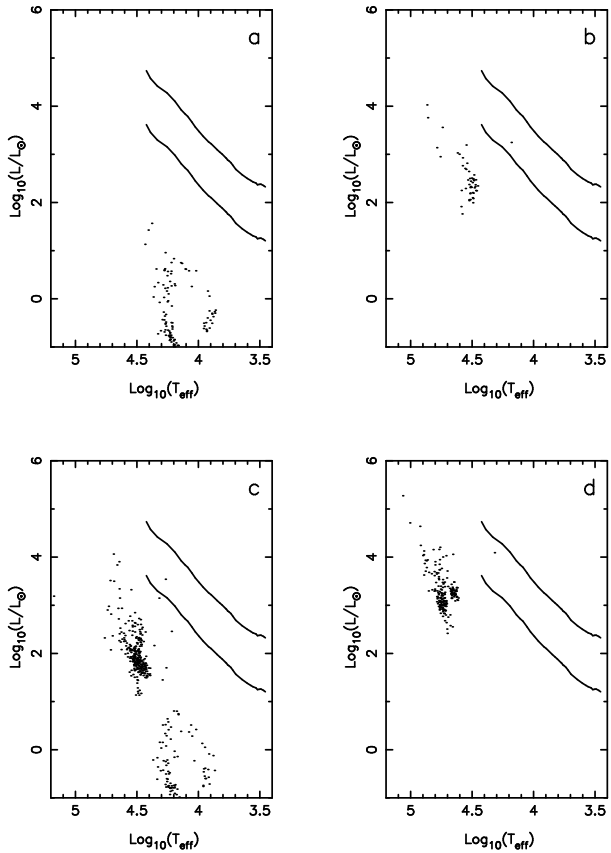,angle=0,width=80mm}
\caption{Snapshots of the HR diagram positions of some of our stripped star populations. Shown are population snapshots at age 
$4 \times 10^{9} {\rm yr}$ assuming 
stripping leaves 1\% of the stellar envelope remaining, with input IMFs with lower mass limits of (a) $0.01 \msun$ , (b) $3 \msun$  with flat 
IMF slope, (c) 
hybrid IMF with 4/5 `normal' IMF, 1/5 $3 \msun$ lower limit with normal IMF slope and (d) $6 \msun$ lower limit. Also shown are 
the approximate luminosity limits corresponding to the K-band magnitudes for 
observed Sgr A$^{*}$ cluster stars, as detailed above; our populations should ideally contain $\sim 20$ stars in the blue region of this area 
(S2 has $T_{eff} \sim 30000 K$, Ghez et al. 2003), not too many above 
the upper line (though the variable extinction may allow the upper line to be increased by a few magnitudes), and no red stars above either 
line. The `normal' IMFs fail to produce anything like a Sgr A$^{*}$ cluster star, whereas the other IMFs produce acceptable 
distributions (some blue stars, no red stars) but at too low a rate.}
\vfil}
\label{fig5}
\end{figure}

In Fig. 6 we show some output models from the late-stripping population, i.e. the population including main-sequence stars which scatter 
onto the giant loss cone and are stripped on a later pass of the black hole when they have evolved to larger radii. 
It is notable that the population is much-increased; the effective 
scattering rate for giant stars is much higher in this scenario. Models which are top-heavy and which have less than 5 \% remaining envelope 
produce reasonable Sgr A$^{*}$ cluster populations.

\begin{figure}
\vbox to140mm{\vfil
\psfig{figure=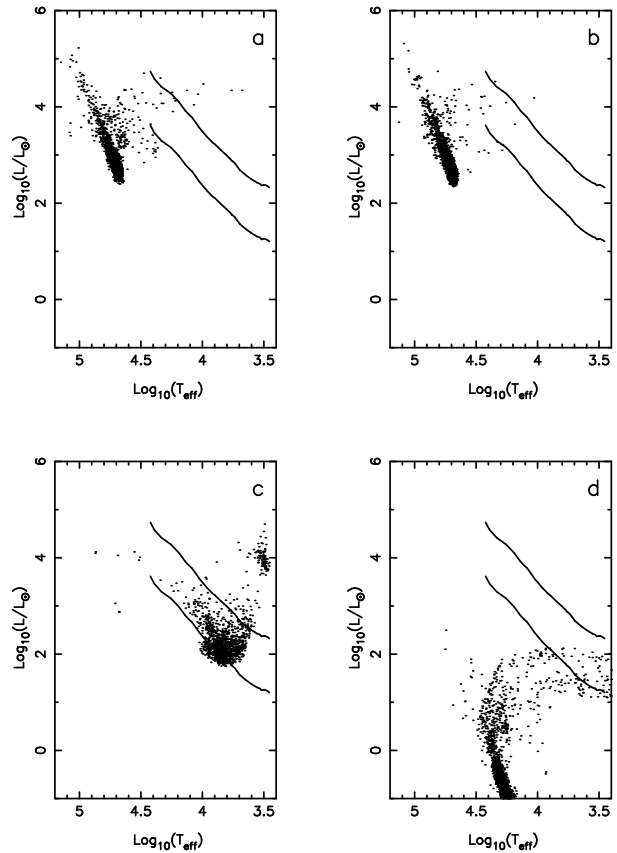,angle=0,width=80mm}
\caption{As Fig. 5, but showing successful and unsuccessful late-stripping populations: (a) Top-heavy ($6 \msun$ limit) IMF with 
1 \% remaining envelope; (b) the same but with 
0 \% remaining envelope; (c) Top-heavy ($3 \msun$ limit) IMF with 5 \% remaining envelope, which produces a 
too-red population; (d) `normal' IMF with 1 \% remaining envelope, which again produces a large red population.}
\vfil}
\label{fig6}
\end{figure}

\subsection{Red Stars}

Some of the models which produce an acceptable population of blue stars also produce a small number of red ones. 
Whilst excluding models which produce a few red stars does not rule every model out, it does rather restrict the 
parameter space over which the models work. 
Of course, we could simply seeing the Galactic Centre at an unusual time when the red star population is low. 
However, the existence and observability of these model red stars may not be guaranteed.

One possible reason that we might not see any red stars in the Galactic Centre, even though we predict some on close orbits, is that 
most such stars are convective and have relatively large radii. Even if they are not overflowing their tidal lobes, the magnitude of 
tidal effects on stars with convective envelopes is much greater than that on stars with radiative envelopes (see e.g. Hurley et al. 2002). 
This arises from the different dissipation mechanisms operating in each case. Stars which are not overflowing their tidal lobes but pass within a
few $r_{\rm tidal}$ of the black hole are still subject to smaller-scale tidal interaction at periastron (Novikov et al. 1992). The likely outcome 
is a decrease in the eccentricity of the orbit, with the orbital energy thus liberated deposited in the internal energy of the star. This in turn 
leads to a radius increase and hence stronger interaction on the next periastron passage. Novikov et al. 
suggest that this could lead to the destruction of the star in a few orbits, with the energy deposited exceeding the total binding energy. Such a 
mechanism could efficiently rid the Sgr A$^{*}$ cluster of its red members. For example, if we remove from our sample of stars those which have 
convective envelopes and pass within $ 3\,r_{\rm tidal}$ at pericentre, this effectively gets rid of all but the least massive red stars.

A further complication is the effect of their extreme environment on their appearance. It has been suggested (Jimenez et al 2006) that
irradiation of giant stars by even a relatively low-luminosity AGN can, via the destruction of molecules in their atmospheres, 
make them appear significantly bluer. They claim such stars would have a similar appearance to the weak-lined Sgr A$^{*}$ cluster population. 
Such a claim should be handled cautiously, since irradiation typically affects predominantly the side of the star facing the radiation 
source, and it is unlikely that all the weak-lined Sgr A$^{*}$ cluster stars are fortuitously presenting their irradiated faces in our direction at 
one moment. It does, however, suggest that even if we produce what should look like giants on Sgr A$^{*}$ cluster star orbits they are not guaranteed 
to appear to be so. Putting into our simulations the level of apparent surface temperature increase suggested by the simulations 
of Jimenez et al. we find a relatively mild effect on the overall populations, certainly not enough to pass off most red stars as 
members of the Sgr A$^{*}$ cluster.

Overall, we find a significantly better fit with top-heavy IMF populations and with smaller amounts of remaining envelope. In particular, if more 
than a small proportion of input stars are of relatively low mass we find too great a population of low-luminosity but still visible red stars, 
which cannot be explained away by either of the above mechanisms. In fact, the absence of such observed stars could place
an upper limit on the scattering rate of stars from outside the central region (see section 5).  Whilst the orbits that we find for these low-mass stars 
are wider and more eccentric than both those of the more massive stars and the observed Sgr A$^{*}$ cluster orbits, implying that they spend most of their lives far 
from pericentre where they may be less observable, we generate 
a large enough number of them in these cases that we would still expect some to be visible. 
That we do not suggests that, if this model for the Sgr A$^{*}$ cluster is to work, the input population is largely drawn 
from the central region itself. For example, late-scattering with a $6 \msun$ lower IMF limit produces a plausible Sgr A$^{*}$ cluster star population (Fig. 7)
 for the scattering rate we use, and in general input IMFs which are lacking in stars below $3 \msun$ can be made to produce plausible Sgr A$^{*}$ cluster populations for scattering rates which are not much different.

\begin{figure*}
\vbox to120mm{\vfil
\psfig{figure=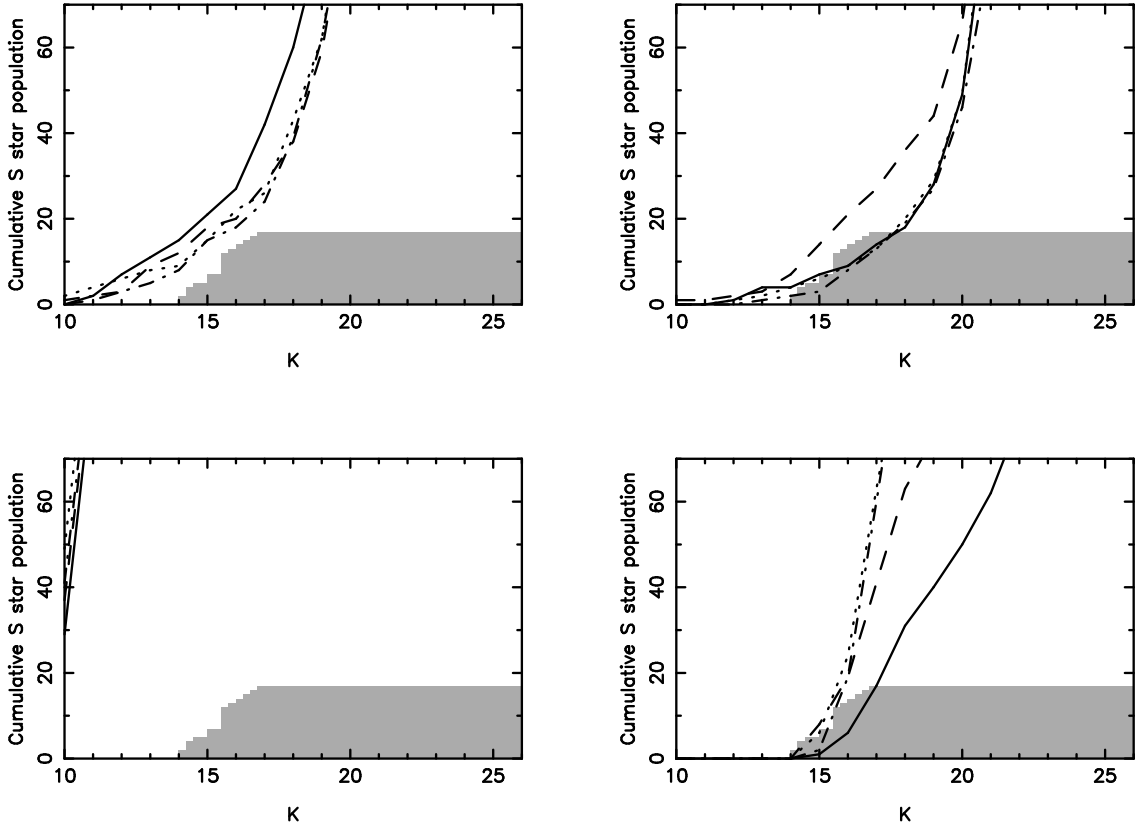,angle=270,width=150mm}
\caption{Cumulative K distribution for the populations shown in Fig. 6. Shown for each input population are snapshot K 
distributions at 1 (solid line), 2 (dashed line), 3 (dot-dashed line) and 4 (dotted line) $\times 10^{9}$ years, against 
the observed distribution (Ghez et al. 2005, in grey).}
\vfil}
\label{fig7}
\end{figure*}

The division by Eisenhauer et al. (2005) and others of the Sgr A$^{*}$ cluster star spectra into two classes suggests the possibility of two separate 
observable populations. There are a number of possible explanations for this within the framework of our model. Firstly, it is 
interesting to note that the simulations of Jimenez et al. (2006) suggest spectra similar to those of the weak-lined 
population may be produced by the irradiation of giants. In some of our simulations we find some giants amongst our Sgr A$^{*}$ cluster stars. It could 
be that, when irradiated, these make up the weak-lined population and the initially-blue stars make up the strong-lined population.
However as noted previously, the sort of temperature increases they predict are not enough to make a significant difference to 
the red stars that we find.
It is also possible, if some of the input stars are supplied by a top-heavy IMF and some are supplied by a `normal' IMF, that we 
are seeing two populations with different input mass distributions. Finally, the evolutionary tracks of many of our more luminous 
stripped stars take them into the K-band visible region on the HR diagram twice: once moving bluewards shortly after stripping, and 
once moving back a little way redwards after a period on the helium main sequence. 

\subsection{Orbits}

For both potential populations, the resulting post-stripping orbits are governed by the fraction of the envelope which is lost in tidal 
disruption, as this determines the energy budget available. Detailed SPH simulations are ongoing to determine the outcome of the stripping process
(Dale et al., in preparation). At present it is an unknown, so we treat the amount of envelope remaining as a free parameter.  
Constraints are introduced, as in the previous section, by the requirement 
that the stripped star population is luminous enough to be observable (difficult if no envelope is left) and mostly blue (difficult if 5\% or more 
of the envelope is left). This may be seen in Fig. 4.

We consider primarily the situations in which 5, 1 and 0\% of the envelope remains. Since a great deal of the envelope binding energy is 
concentrated in the region nearest the core, these scenarios have significant differences in available energy budget. In particular, for stars with 
5 \% or more of the envelope remaining, the relatively low amount of energy extracted leads to a distribution of orbits which are rather weakly-bound. 
As the periastron distance is constrained by the need to initially pass close to the central black hole, this leads to very high typical eccentricities 
-- 0.99 or more. The same applies to populations with a large proportion of low-mass stars, which also have smaller envelope binding energies.
This includes the `normal'-IMF population. These highly-eccentric orbits provide a poor match to the sample of Sgr A$^{*}$ cluster orbits 
which are known (see e.g. table 2 of Eisenhauer et al. 2005). Whilst the Sgr A$^{*}$ cluster orbits are eccentric, their measured 
eccentricities range from 0.36 to 0.94. However, the models which produce unrealistic orbits are also those which have been previously ruled out for 
producing too many red stars. Stars from top-heavy IMFs and those which have more envelope removed, favoured as Sgr A$^{*}$ cluster star 
progenitors in the previous section, also provide the best match here.
The greater the amount of envelope removed, the better the fit we find to 
observed orbits. Shown in Fig. 8 are snapshot orbital distributions for some of our models which produce plausible luminosity distributions.

\begin{figure*}
\vbox to125mm{\vfil
\psfig{figure=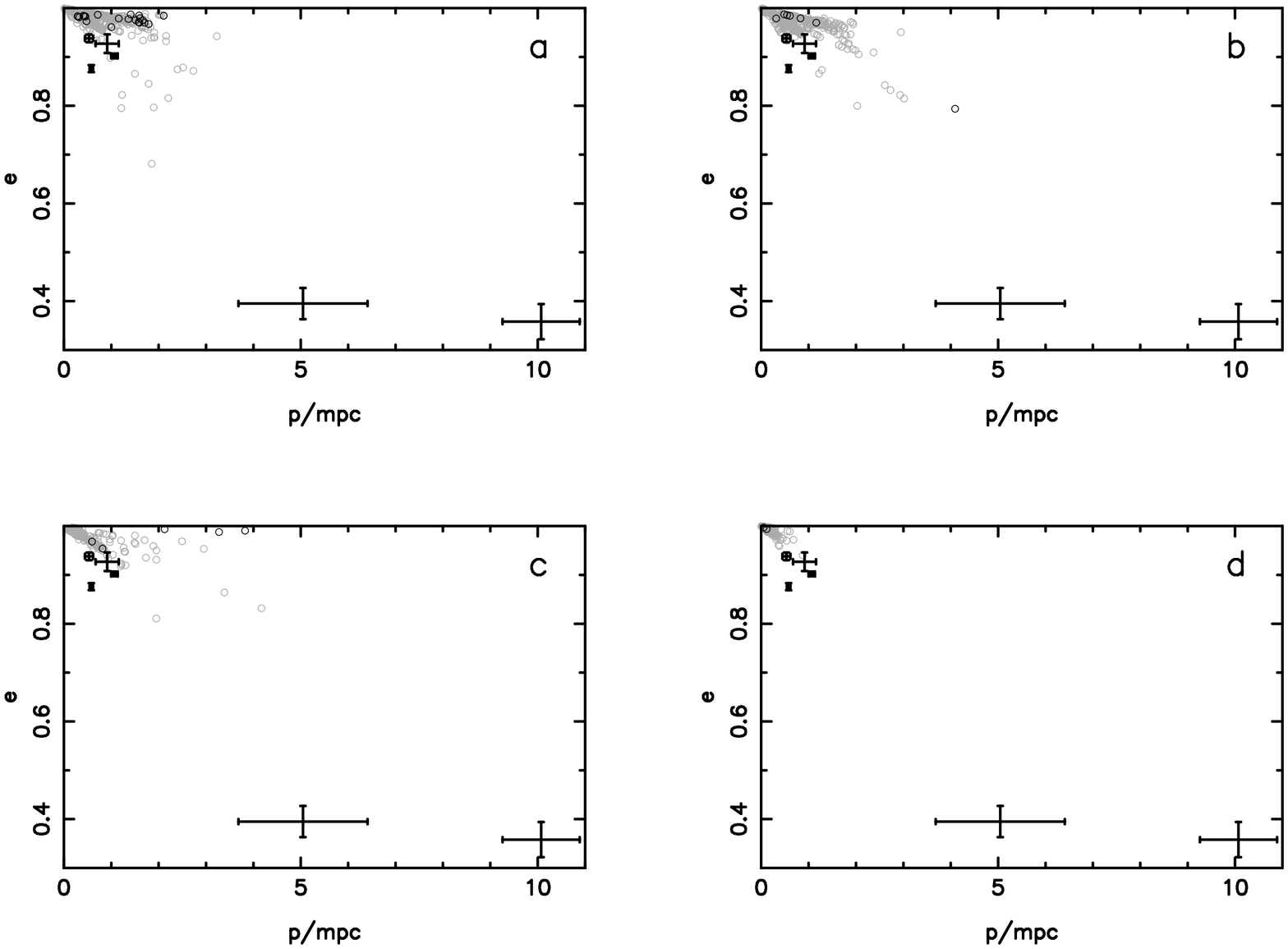,angle=270,width=150mm}
\caption{Snapshot of Sgr A$^{*}$ cluster orbits at $4 \times 10^{9}$ years, for (a) late-stripping IMF with Salpeter slope and lower limit of $6 \msun$, 
1 \% remaining envelope; (b) the same but with 0 \% remaining envelope; (c) Arches-type IMF with 1 \% remaining envelope; (d) First-pass stripping 
with lower IMF limit of $3 \msun$ and 1\% remaining envelope. Black points indicate visible, grey points non-visible Sgr A$^{*}$ cluster stars. Note that in general 
the visible stars are spread relatively evenly amongst the distribution of orbits. Shown also are the six current observed Sgr A$^{*}$ cluster orbits 
(Eisenhauer et al. 2005).}
\vfil}
\label{fig8}
\end{figure*}

Although the parameters of three of the six Sgr A$^{*}$ cluster stars which have detailed orbital solutions (S12, S14 and S8) are easy to match with observable models, 
the star with the best-determined 
orbit, S2, is slightly outside the area in eccentricity and periastron occupied by our models, and the two relatively low-eccentricity stars, S1 and S13, 
are well outside this region. This suggests later orbital evolution after the initial stripping event.
As discussed in Davies \& King (2005), such orbits could be produced by scattering interactions with 
other stars in the central region after the initial stripping event -- a not entirely surprising happening since, as 
we have already discussed, the relaxation time in the central region needs to be small for the model to work in the first place.
The relaxation time in the central arcsecond is likely to be less than $10^{8}$ yr (Sch{\"o}del et al. 2003), perhaps as low as $10^{7}$ yr 
(Davies \& King 2005) given the large number of compact remnants (white dwarfs, plus some neutron stars and black holes) which are likely to be there. 
Typically, many of the cluster stars which are observable in the top-heavy IMF model live around 1 -- $2 \times 10^{7}$ yr after 
stripping. The portion of the tracks for which 
the star is visible varies from star to star but, whilst some stars are visible for a period of time just after stripping, a substantial 
portion of the visible time generally occurs near the end of the stripped star's life. Thus it is plausible that some of the Sgr A$^{*}$ 
cluster stars have undergone one post-stripping scattering event.

Another possibility is that, even though the Sgr A$^{*}$ cluster stars do not overflow their tidal lobes at pericentre, 
smaller-scale tidal effects are still occurring, 
leading to a slow circularisation of the orbit similar to orbital circularisation in massive binaries. 

\subsection{Rotation}

The Sgr A$^{*}$ cluster stars have been observed to have rotation rates typical of main-sequence B stars (Eisenhauer et al. 2005). In our scenario for 
their origin, the rotation rate depends largely on the details of the envelope removal process. Since we are here
treating that process as a 'black box', it is not possible to comment too deeply on the rotational rates. However, it is likely that 
the stripping process will induce at least some rotational velocity change in the star. Diener et al. (1995) find the effect of 
tidal interaction on their initially non-rotating model stars is to produce both rotation and oscillation. 

\subsection{Further Population constraints}
Besides producing Sgr A$^{*}$ cluster stars like those observed, there are a number of other common-sense constraints on which 
populations are acceptable.

\subsubsection{There should not be too massive a collection of remnants orbiting the black hole}
One of the chief complaints of Goodman \& Paczynski (2005) regarding an AGB star origin for the Sgr A$^{*}$ cluster stars is that a sufficiently large 
population of stripped AGB stars (which only remain in the observable region for a short time) to reproduce observations would 
lead to an unreasonably large mass of non-visible remnants orbiting the black hole (white dwarfs, neutron stars and some black holes, 
depending on the IMF). Whilst the stripped RGB and horizontal branch stars which make 
up the bulk of the Sgr A$^{*}$ cluster in our scenario remain visible for a longer time, this is still potentially a 
problem. Several factors affect the mass of remnants in the central region. Firstly, the amount of mass which enters the region has a strong effect. 
Top-heavy IMF models generally leave more massive remnants, so even if the number of stars coming in is the same the total remnant mass is greater.
For example, `normal' first-pass IMF models leave a few hundred to a few thousand solar masses of orbiting remnants, generally towards the lower 
end of this range. This is because most stars 
which encounter the black hole are disrupted entirely. For top-heavy IMFs the remnant mass can be closer to a few $10^{5}$ solar masses. However 
the total number of remnants is similar to that for `normal' IMFs; the difference is that each remnant is more massive.
 
Second, the mass of orbiting remnants is greater for the late-stripping populations. This is unsurprising, since a much larger proportion of
them encounter the black hole as giants and so avoid complete shredding. For late-stripping top-heavy IMF populations the remnant mass 
resulting can be 
particularly unrealistic and can approach the mass of Sgr A$^*$ itself in the most extreme case. This is problematic, since the late-stripping 
top-heavy IMF models provide the best match to the K luminosity and orbital characteristics. However, the remnant mass is dependent on the 
scattering rate. 
For Salpeter-slope IMFs with a high lower mass limit, the scattering rate is not very different from the $1 \msun, 1 \rsun$ scattering rate; for 
IMFs which are top-heavy because they have a flatter upper slope, such as our Arches-like IMF, the scattering rate is much lower and acceptable 
populations can still be obtained with reasonably small remnant masses. In addition, the above analysis assumes the loss cone is kept 
full -- probably an unlikely state. If instead the scattering rate is constrained by the refilling rate of the loss cone the rate of 
disruptions may be much lower and hence the remnant mass still reasonable, although this will also reduce the number of visible stars unless 
formation happens in bursts, perhaps related to recent star formation in the GC region. However, this latter scenario would have 
trouble explaining the current low luminosity of Sgr A$^{*}$. 

More importantly, the above analysis of the remnant mass assumes nothing happens on the formation of a remnant and no further interaction 
between remnants takes place 
after formation. In the top-heavy IMF case, much of the mass of remnants is made up out of neutron stars and even some black holes. This requires type 
II (potentially IIb or even Ib, since only a small amount of envelope remains) supernovae. It is generally thought (e.g. Lyne \& Lorimer 1994) that 
the formation of a neutron star or black hole by a supernova is accompanied by a large velocity kick arising from asymmetries either in the mass 
ejection or the neutrino emission. Therefore much of the remnant mass could in fact have been ejected from the central region, either inward (i.e. 
the remnant is propelled onto an unstable orbit, from whence it is swallowed) or outward\footnote{Note that the velocity dispersion in the 
central region is rather large; orbital speeds may be $> 1000 {\rm km\,s^{-1}}$. The average kick value from the study of Lyne \& Lorimer is 
$450 {\rm km \,s^{-1}}$. So this mechanism probably accounts for only a portion of the remnants.}. If all such remnants are 
ejected, the total mass of orbiting remnants in the top-heavy late-stripping  models after $10^{10}$ years decreases to 
$\sim 10^{5} \msun$. For non-expelled remnants, interactions arising from the high density of the region may also lead to ejections. In particular, 
feedback interactions may increase the number of ejections, with temporarily-overdense regions having high ejection rates; and, if there is a 
population of massive black holes which have migrated to the central parsec by dynamical friction (Miralda-Escud{\'e} \& Gould 2000) these 
will tend to eject the individually less-massive remnants from stripped-star evolution. If mergers 
between the central black hole and other infalling SMBHs from Galaxy mergers happen periodically (Yu \& Tremaine 2003), the central region 
is likely to be `cleared-out' of its population of remnants when this occurs: in this case, the central remnant population may be only the 
much smaller number of compact objects formed since the most recent SMBH merger.

The number of remnants is also partially dependent on the amount of envelope which the stripping process typically removes from a star. 
The greater the amount of envelope the star retains, the more likely it is to undergo a second expansion phase after central helium burning and 
end up as a helium giant. In this state it is likely to overflow its tidal lobe a second time or at least undergo strong 
tidal interactions (section 3.1). Several outcomes are possible for this second stripping process. If it behaves similarly to the first, 
one would expect a more circular orbit with the same pericentre distance to result. More probably, a second large input of energy and/or repeated tidal 
interactions at periastron may disrupt the star completely (Novikov et al. 1992). In either case, the total mass of remnants is reduced.

\subsubsection{The flaring rate should be reasonable}
The disruption of a star and subsequent accretion of its matter can produce a strong X-ray flare, such as have been observed in 
some AGN (e.g. Piro et al. 1988; Brandt, Pounds \& Fink 1995; Grupe, Thomas \& Leighly 1999; Donley et al. 2002). However, the 
time-scale of the tidal disruption event determines its observability, since 
very long timescale events are not detectable as flares. Red giants with large radii are amongst the stars which have a long 
disruption timescale (Syer \& Ulmer 1999). Therefore a larger disruption rate in which the increase is primarily of giant stars only, such 
as we find in the late-stripping case, does not necessarily result in a greater number of observable flares. Much of the X-ray luminosity 
function of active galactic nuclei may be powered by this method, particularly at the low-luminosity end (Milosavljevi{\'c}, Merritt \& Ho 2006).
Tidal disruption rates have frequently been calculated in the past with a view to working out the rates of X-ray flares (e.g. Rees 1988; 
Magorrian \& Tremaine 1999). More frequent minor flares have also been observed in Sgr A$^*$, which could be the signatures of stars passing 
through the black hole's accretion disk (Nayakshin, Cuadra \& Sunyaev 2004).

Even for our most extreme late-scattering models, the rate of tidal disruption is never more than a factor of 4.5 greater than for the case 
in which all stars are $1 \msun$, $1 \rsun$ and are stripped on their first pass. This corresponds to a maximum rate for 
the Galactic Centre of around $2 \times 10^{-3} {\rm disruptions \, yr^{-1}}$. For realistic models which provide a reasonable fit to the 
observational data and to the other constraints discussed here the rates of tidal disruption are generally lower. For example, the total rate 
of tidal disruptions in the case of the `Arches' IMF is only 20\% of that for the $1 \msun$, $1 \rsun$, first-pass stripping case.
However, as yet the constraints on flaring rates are relatively weak. Donley et al. (2002) find the rates to be approximately 
$9 \times 10^{-6} {\rm galaxy^{-1} \, yr^{-1}}$ for inactive and $8.5 \times 10^{-4} {\rm galaxy^{-1} \, yr^{-1}}$ for active galaxies. Our 
predicted flaring rates are nearly all comfortably in this range. This is therefore also a rather weak constraint on the models when 
compared to the others. 

A further observational limit which has the potential to constrain the rate of inspirals into the black hole in the future is their gravitational wave 
emission (e.g. Hopman \& Alexander 2006). 

\subsubsection{The accretion rate should be reasonable}
The tidal disruption process is also a method of black hole feeding, and it is likely that at least some fraction of the 
stripped matter (perhaps around half; Rees 1988) is accreted onto the black hole. Although the current low luminosity of Sgr A$^{*}$ 
indicates that its accretion rate at the moment is much smaller than expected (e.g. Baganoff et al. 2003; Zhao, Bower \& Goss 2001), 
it must obviously have accreted enough matter in the past in some fashion to grow to its 
current mass. Tidal shredding is by no means the only method of feeding a black hole; at least some of the mass is likely to 
arise from gas accretion (e.g. Wada 2004). If the amount of matter accreted over $10^{10}$ years from tidal disruptions is significantly over 
$3.6\times 10^{6} \msun$, this suggests 
the model is unrealistic. Of course, the amount of accretable matter from loss cone stars changes over the lifetime of the black hole, as the 
scattering rate depends on the black hole mass. Wang \& Merritt (2004) find, perhaps counterintuitively, a wider loss cone and thus greater 
scattering rate onto lower-mass black holes. The fraction of matter which is accreted is also a matter of debate and may be rather lower than half, 
perhaps closer to 10\% (Ayal, Livio \& Piran 2000).
Late-stripping models in general have larger amounts of accreted mass simply because more stars are stripped. Similarly top-heavy IMF models 
have larger amounts of accreted mass because the total amount of mass in the stars encountering the black hole is greater. Generally the models 
which can be rejected on these grounds are the same ones which can be rejected on the grounds of having too great a mass of remnants, therefore.
Top-heavy late-stripping models can produce produce acceptable levels of accretion if the scattering rate is not too high (this occurs with flatter 
upper IMF slopes, for example, so the Arches-like model is again good here) or if the fraction of matter accreted is closer to 
the Ayal et al. (2000) value than to a half.

\section{Discussion}

If the Sgr A$^{*}$ cluster stars arise from relatively low mass progenitors, the `paradox of youth' (Ghez et al. 2003) is no longer a problem. 
However, it seems apparent from our models that the requirement for young stars is only slightly relieved; 
the model which matches most closely to the HR diagram position of a $15 \msun$ ZAMS B star, for example, is an initially 
$9 \msun$ star which undergoes stripping at an age of 29.5 Myr to remove all but about $2 \msun$ of its mass.
But the $15 \msun$ star itself stays on the main sequence for 10 Myr. If the Sgr A$^{*}$ cluster stars are produced in 
this manner, then, it is likely that the disks of apparently massive young stars within a parsec of Sgr A$^*$ which are the next 
furthest objects out from them were in fact formed in place (Nayakshin \& Sunyaev 2005) and that the nearby Galactic Centre region in general 
supplies many of the scattered stars which become visible on Sgr A$^{*}$ cluster orbits after stripping. 

    The IMF of the scattered stars is the most important ingredient in producing a plausible Sgr A$^{*}$ cluster model from stripped stars. Most 
top-heavy IMF models can be made to produce enough visible blue stars and few enough red stars with scattering rates which are in the range 
of those claimed in the literature.
However, as discussed above, other constraints are important. In particular, the fraction of the 
matter produced by stripping stars which is accreted onto the black hole should be relatively low -- a quarter or less -- if this model
is not to produce a greater mass for Sgr A$^{*}$ than is observed, and the relaxation time in the central $0.7$ pc needs to be below 
$10^{8}$ yr. The most successful model when all these constraints are taken into account is the late-stripping `Arches-like' model.
If the constraints are not met, it is probable that this mechanism still operates at least at some level, and could therefore 
be responsible for some of the Sgr A$^{*}$ cluster stars.

   The generally poor results we find with `normal' IMFs may also provide an interesting constraint on the tidal disruption rate from regions 
outside the central cusp. Stars scattered from these regions are likely to belong to such an IMF -- in fact, it is likely that they are entirely 
deficient in high-mass stars due to the short lifetimes involved. Even though lower-mass stars end up on higher-eccentricity orbits, in the 
late-stripping case enough red stars are made that we would still expect a population of red stars observable amongst the Sgr A$^{*}$ cluster stars. The fact 
that we do not see such a population potentially suggests limitations either on the tidal stripping rate from a full loss cone or the rate stars 
are scattered onto the loss cone. In Fig. 8 we plot the number of visible red stars likely with radius as a function of tidal disruption rate. A 
significant red population arises from scattering rates above about $10^{-4} {\rm disruptions\, yr^{-1}}$.

\begin{figure}
\vbox to70mm{\vfil
\psfig{figure=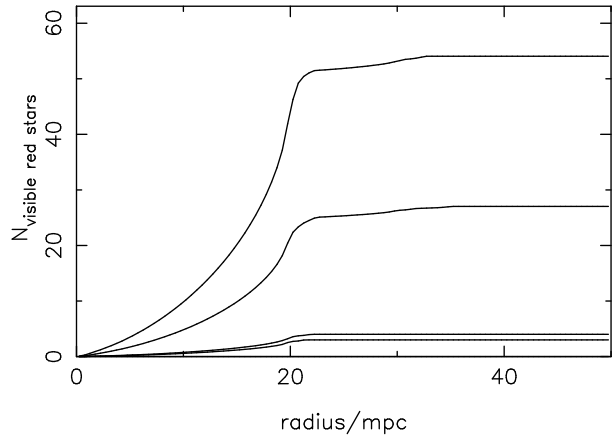,angle=270,width=80mm}
\caption{Cumulative number of visible red stars with radius expected in the Galactic centre from `normal'-IMF late-scattered stars 
after $4 \times 10^{9}$ years, with 1 \% remaining envelope 
and tidal disruption rates (upper line to lower line) $10^{-3}$, $5 \times 10^{-4}$,$10^{-4}$
and $10^{-4} {\rm disruptions\, yr^{-1}}$.}
\vfil}
\label{fig9}
\end{figure}

\subsection{Observational Prospects}

  Observationally, there are a number of ways that it may be possible in the future to distinguish between this scenario and others for the 
formation of the Sgr A$^{*}$ cluster. An unavoidable part of the stripped-star scenario is that there is a large population of single compact 
objects, many of which will be neutron stars or even black holes, in the central region. This extra dark mass should 
cause the orbits of the cluster stars to precess slightly (Mouawad et al. 2005; Weinberg, Milosavljevi{\'c} \& Ghez 2005). 
Whilst current telescopes are unable to measure the stellar orbits with enough accuracy to make out this effect, future 30 -- 100 m 
Extremely Large Telescopes (ELTs) should be able to effectively weigh the extended matter distribution around the black hole using this method
(Weinberg et al. 2005). There is also the possibility of observing these remnants directly. Pulsars orbiting Sgr A$^{*}$ should be visible with 
SKA (Cordes et al. 2005). Recent work (Wu et al. 2006; Deegan \& Nayakshin 2006) also suggests that 
single neutron stars and black holes may be observable in the Galactic Centre region if they can accrete enough 
gas from the local ISM to be X-ray bright. This effect has been suggested as an explanation for the overdensity of X-ray point sources
towards the Galactic centre (Muno et al. 2003). However, although the absence of a cusp of compact objects would disprove the 
stripped-stars theory, the presence of a such a cusp is not in itself definitive proof as 
massive compact objects (5 -- $10 \msun$ black holes) are expected to migrate to the central parsec from outside via 
dynamical friction (Miralda-Escud{\'e} \& Gould 2000), and such objects may be efficient at ejecting lower-mass stellar 
remnants from the area.

With future space-based or AO telescopes performing deeper observations in the central parsec region, there exists the 
possibility of constraining the initial or at least current mass function in this area. As is apparent above, the stripped-stars model 
only works if the local IMF is top-heavy. In particular, the best-fit IMF is `Arches-like', with a relatively flat upper slope.
There are already limits on the size of any potential low-mass star population from X-ray emission (Nayakshin \& Sunyaev 2005), 
and the extremely high WR and LBV population in the central parsec in comparison to other massive clusters with a similar 
O and B star population (e.g. Tanner et al 2006) suggests that the upper IMF slope may be flatter than normal. In our simulations we
find a distribution of cluster star luminosities which extends down below the boundary of what is currently observable, so we would
expect future deeper observations of the Sgr A$^{*}$ cluster itself to reveal more blue stars on similar orbits, as indicated in figure 7.

Finally, there are a number of properties of the cluster stars themselves which could be used to confirm or deny different theories 
as to their origin. Whilst the stripped cores we have produced appear similar to main-sequence B stars in some observational respects, they should 
be by no means identical. In particular, their surface compositions will differ. If the Sgr A$^{*}$ cluster stars 
are B stars then their surface H and He abundances should be typical of the metallicity of the Galactic centre region -- i.e. about solar 
(e.g. Najarro et al. 2004). If they are stripped stars, then they should be somewhat H-depleted, probably with a surface mass fraction of 
hydrogen between $0.6$ and $0.1$, and these surface abundances should vary strongly from star to star. As the orbit is dependent on the 
amount of envelope removed, there may be correlations between the periastron and eccentricity and the luminosity of the Sgr A$^{*}$ 
cluster stars which could be investigated in the larger sample of stars and orbital solutions available from deeper observations. Here 
we would expect a slight correlation between high luminosity and lowish eccentricity -- however, this may be wiped out entirely if 
further scattering events take place to form the low-eccentricity orbits (section 4.2).
A much more definite test would be if we were lucky enough to observe a supernova amongst these stars, 
since we expect some of our top-heavy IMF models to undergo type IIb supernovae. B stars on the same orbits probably avoid supernovae; once 
they evolve to larger radii, it seems probable that they are completely disrupted by tidal interactions with the hole. 

\subsection{Central Star Clusters in Other Galaxies}
It is interesting to note that the mass of the central black hole is critical in this picture in deciding 
whether a central cluster like the one observed can be formed and what its characteristics are. There 
have already been observations suggestive of central clusters of young stars in other galaxies (e.g. Lauer et al. 1998), 
and it is likely that if the stripped-stars mechanism operates in our own Galactic centre it also operates in others. 
This raises the possibility of observing clusters analogous to the Sgr A$^{*}$ one around other SMBHs, and the question 
of what properties such clusters would have. Firstly, the scattering rate depends 
on $m_{6,BH}$ (where $m_{6,BH}$ is the black hole mass in units of $10^{6} \msun$) roughly as $m_{6,BH}^{-0.25}$ 
(Merritt \& Wang 2004), i.e. a black hole of a few $10^{8} \msun$ should, discarding any other effects, have a Sgr A$^{*}$-like cluster 
only one-third as numerous as the Galactic one. Secondly, the Schwarzschild radius is greater for more massive black holes. 
For black hole masses 
comparable to that of Sgr A$^{*}$ the Schwarzschild radius is only a few solar radii and most solar-type stars passing 
within a tidal radius of the black hole are disrupted rather that directly swallowed. However the ratio of 
Schwarzschild radius to tidal radius is approximately 
\begin{equation}
\frac{R_{Sch}}{R_{T}} \sim 2.1 \times 10^{-2} \, \frac{m_{*} m_{6,bh}^{2}}{r_{*}}
\end{equation}
where $m_{*}$ and $r_{*}$ are the mass and radius of the scattered star in solar units 
(e.g. Beloborodov et al. 1992, Ivanov \& Chernyakova 2006).
Black holes of a few $10^{8} \msun$ will therefore swallow solar-type stars directly rather than tidally 
disrupting them and accreting some of the gas. Many of the visible Sgr A$^{*}$ cluster stars in our first-pass models arise from 
horizontal branch (HB) stars, as this is the longest phase during which the star has a well-defined core. As HB stars are also the most 
compact of the potential progenitor stars, they will be the first section of the progenitor population to be swallowed rather than 
stripped as black hole mass increases. Taking typical
values of mass and radius for HB stars which are visible as Sgr A$^{*}$-like cluster stars when stripped suggests that the 
black hole mass above 
which the Sgr A$^{*}$-like cluster population drops dramatically should be around $10^{10} \msun$ -- i.e. this effect will be very small 
in most cases. However, the fate of lower-mass stars and thereby potentially the fraction of gas from scattered stars 
which accretes should be significantly different at the upper and lower ends of the known SMBH mass spectrum.
In particular,  whilst the scattering rate may be lower for more massive black holes (Wang \& Merritt 2004) 
there is a mild compensatory effect in terms of extra mass accretion. 
The observation of clusters obeying the mass relation above in other galaxies would act as significant evidence in favour of this 
method of forming the Sgr A$^{*}$ cluster.

The effect of black hole mass on scattering rate is also important if tidal stripping is to be considered as a general 
AGN fuelling mechanism (Milosavljec, Merritt \& Ho 2006). As noted above, the amount of mass supplied in this way is 
certainly sufficient for such a process. The decrease in loss cone angle as the black hole mass increases suggests that 
in this picture it is the early life of the massive black hole which is characterised by more rapid accretion and that the 
growth of supermassive black holes gradually slows down with time. However, the evolution of massive black holes is 
unlikely to be this simple. If galaxy mergers are a frequent event then there are likely to be phases in which the black holes
at the centres of some galaxies are binaries. Such a situation can cause a large increase in the tidal disruption rate 
(Ivanov, Polnarev \& Saha 2005) and will also have a strong disruptive effect on any pre-existing Sgr A$^{*}$-like cluster, with a high ejection 
rate from the central region (Yu \& Tremaine 2003).

\section*{Acknowledgements} 

We would like to thank Sergei Nayakshin, Walter Dehnen, 
Hagai Perets and Mark Freitag for their very useful and helpful comments.
LMD is supported by the Leicester PPARC rolling grant for
theoretical astrophysics. ARK gratefully
acknowledges a Royal Society Wolfson Research Merit Award.
MBD is a Royal Swedish Academy Research Fellow supported by a grant
from the Knut and Alice Wallenberg Foundation.

{\vspace{0.5cm}\small\noindent This paper
has been typeset from a \TeX / \LaTeX\ file prepared by the author.}

\label{lastpage}

\end{document}